\documentclass[12pt]{iopart} 
\usepackage{epsfig}                   

\begin{document} 
\begin{flushright}
 PITHA 00/29\\
 TPR-00-25\\
 hep-ph/0012222\\
 18 December 2000\\
\end{flushright}
\title[Lifetime difference of $B_s$ mesons - Theory status]{Lifetime 
       difference of $B_s$ mesons - Theory status\footnote{
Talk presented at the UK Phenomenology Workshop on Heavy Flavour 
and CP Violation, 17 - 22 September 2000, St John's College, Durham, 
proceedings to appear in J. Phys. G.}}

\author{Martin Beneke}
\address{Institut f\"ur Theoretische Physik E, RWTH Aachen, Sommerfeldstr. 28,
         D-52074 Aachen, Germany}
\ead{mbeneke@physik.rwth-aachen.de}

\author{Alexander Lenz}
\address{Fakult\"at f\"ur Physik, Universit{\"a}t Regensburg, 
         D-93040 Regensburg, Germany}
\ead{alexander.lenz@physik.uni-regensburg.de}

\begin{abstract}
         We give an update on the lifetime difference of $B_s$ mesons 
         which accounts for recent lattice bag parameter results,
         and obtain $(\Delta\Gamma/\Gamma)_{B_s}=(9.3^{+3.4}_{-4.6})\%$. 
         We then review the current theoretical un\-cer\-tain\-ties
         and conclude 
         with a pessimistic perspective on further improvements.
\end{abstract}
\section{Status of theory}
      The width difference in the $B_s$ system is expected to
      be the largest rate difference in the B hadron sector \cite{largest}.
      Some experimental bounds on this quantity from
      LEP \cite{dgexplep} and CDF \cite{dgexpcdf} exist, 
      and in the near future
      $\Delta \Gamma_{B_s}$ will be measured quite precisely \cite{dgexpcdf}.
      Several factors contribute to the interest in $\Delta \Gamma_{B_s}$:
      a large value of the width difference opens up the possibility for
      novel studies of CP violation without the need for tagging
      \cite{notagging}.
      Moreover, an experimental value of $\Delta \Gamma_{B_s}$ would give
      information about the mass difference in the $B_s$ system 
      \cite{dgdm} (although at this moment it appears that the mass 
      difference will be measured sooner than the width difference). 
      Another interesting point is 
      that new physics can only lead to a decrease of the width difference
      compared to the standard model value \cite{grossman}. An experimental 
      number which is
      considerably smaller than the theoretical lower bound, would
      thus be a hint for new physics that affects $B_s$-$\bar{B}_s$ mixing.
      Besides the need for a reliable theoretical prediction of 
      $\Delta \Gamma_s$ in order to fulfill the above physics program 
      it is of conceptual interest to compare experiment and theory in order
      to test local quark-hadron duality, which is the underlying assumption 
      in calculating heavy quark decay rates. 
      One can show  that duality holds exactly in the limit 
      $\Lambda_{\rm QCD} \ll m_b - 2 m_c \ll m_b$ and $ N_c \to \infty$ 
      \cite{aleksan}. So far no deviation from duality has been 
      conclusively demonstrated experimentally and theoretical 
      models of duality violation in $B$ decays tend to predict 
      rather small effects \cite{uraltsev}.

During the last year there has been remarkable interest in the 
lifetime difference of $B_s$ mesons. Besides several experimental studies, 
many lattice calculations of the relevant nonperturbative constants 
were done \cite{dglajap}, \cite{dglait}, \cite{dglaes}.
This improvement in theory input motivates the present update of the result 
presented in \cite{dgbsnlo}. We also clarify the origin of seemingly 
disagreeing recent evaluations of $\Delta \Gamma_{B_s}$.

The theoretical status of the lifetime difference computation is 
as follows: the heavy quark expansion (HQE) allows us to 
expand the decay rate of heavy mesons in inverse
powers of the heavy quark mass.
The leading term is described by the decay of a free quark (the so-called 
parton model) and therefore equal for all $B$ hadrons.
The next term is suppressed by two powers of the heavy quark mass; 
it is related to the kinetic and the chromomagnetic operator. 
The so-called weak annihilation and 
Pauli interference diagrams contribute first at third order. 
Operators involving the spectator quark begin to appear at this 
order and these are mainly responsible for the lifetime 
differences of $B$ hadrons. (A smaller contribution comes from the 
matrix elements of the kinetic and chromomagnetic operator; for 
the lifetime difference of $B_s$ mesons these terms vanish 
identically.) We can write for the decay rate 
difference of two mesons
\begin{displaymath}
\Delta \Gamma = \frac{\Lambda^3}{m_b^3} \left[
\left( \Gamma_3^{(0)} + \frac{\alpha_s}{4 \pi} \Gamma_3^{(1)} + \dots 
\right) +
\frac{\Lambda}{m_b} \left(\Gamma_4^{(0)} +\dots + \right) + \dots
\right] \; .
\end{displaymath}
The $\Gamma_i$'s are products of perturbatively
calculable Wilson coefficients  and matrix elements, which have to be 
determined by some  non-perturbative methods like lattice-QCD or sum rules. 
For a comparison of experiment and theory we need besides the experimental
value several theoretical ingredients.
First, the perturbative prediction in leading order ($\Gamma_3^{(0)}$), 
second, corrections due to the strong interaction ($\Gamma_3^{(1)}:$ NLO-QCD), 
third, subleading $1/m_b$-corrections ($\Gamma_4^{(0)}$)
and last, but not least, the determination of the appearing non-perturbative 
parameters.
For $(\Delta \Gamma/\Gamma)_{B_s}$ we have all these pieces:
the leading term was calculated in \cite{dgbslo} in the factorization 
approximation; 
$1/m_b$-corrections were determined in \cite{bbd}; 
the $\alpha_s$-corrections are given in \cite{dgbsnlo}; 
and the lattice values for the decay constant $f_{B_s}$ and the bag 
parameters $B$ can be taken from computations directed at the mass 
difference. $B_S$, which is specific to the width difference was 
calculated in \cite{dglajap}, \cite{dglait}, \cite{dglaes}.
(The matrix elements of the dimension-7-operators, which emerge in
$\Gamma_4$, have been estimated up to now only 
in vacuum insertion approximation \cite{bbd}.)
As the calculation of $\Gamma_3^{(1)}$ 
was the first to consider QCD corrections to spectator effects, 
there is also a conceptual point of interest. Soft gluon emission
from the spectator $s$ quark leads to power-like infrared divergences, 
which would spoil the HQE. In \cite{dgbsnlo} the infrared safety of
$\Delta \Gamma_s$ was explicitly shown, in agreement with the 
theoretical argument of \cite{bigi}.

Compared to $\Delta \Gamma_s$ our knowledge about the lifetimes ratios 
$\tau (B^+)/\tau (B_d)$ and $\tau (\Lambda_b)/\tau (B_d)$ is much poorer. 
Here only $\Gamma_3^{(0)}$ is known
\cite{lifetime}, while the calculation of $\Gamma_3^{(1)}$ and 
$\Gamma_4^{(0)}$ is still missing. Moreover we only have preliminary 
lattice studies for the bag parameters in $\Gamma_3$ \cite{durhamflynn}.
Results from sum rules for the bag parameters are discussed in 
\cite{durhamdefazio}. 
For these lifetime ratios more work needs to be done.

\section{Numerical update and uncertainties}
\label{sec:update}

The following update is based on the NLO expressions given in 
\cite{dgbsnlo}. The width difference is normalized to the
semi-leptonic $B_s$ branching fraction; numerically we find 
\begin{equation}
\label{numform}
\hspace*{0cm}
\left(\frac{\Delta \Gamma}{\Gamma}\right)_{B_s} = 
\left(\frac{f_{B_s}}{230\,\mbox{MeV}}\right)^{\!2}\left[
0.007 B(m_b)+0.132
\,\frac{M_{B_s}^2 B_S(m_b)}{(\bar{m}_b+\bar{m}_s)^2} - 
0.078\right]
\end{equation}
We factored out the decay constant and use $B$ and $B_S$ to 
parametrize the matrix elements
\begin{eqnarray}
\label{qb}
\langle\bar B_s|(\bar b_is_i)_{V-A}(\bar b_js_j)_{V-A}|B_s\rangle 
&=& \frac{8}{3}f^2_{B_s}M^2_{B_s} B, \\
\label{qsbs}
\langle\bar B_s|(\bar b_is_i)_{S-P}(\bar b_js_j)_{S-P}|B_s\rangle 
&=& -\frac{5}{3}f^2_{B_s}M^2_{B_s}
\frac{M^2_{B_s}}{(\bar m_b+\bar m_s)^2} B_S. 
\end{eqnarray}
Here $\bar{m}_q$ denote $\overline{\rm MS}$ quark masses at the 
scale $m_b=4.8\,$GeV. The third term in square brackets in 
Eq.~(\ref{numform}) is an
estimate of the $1/m_b$ correction in the factorization 
approximation \cite{bbd}. (This correction is 
slightly larger than in \cite{dgbsnlo}, because we now use
$\bar{m}_s=0.1\,$GeV. We will also use $\bar{m}_b=4.2\,$GeV.) We note that
\begin{itemize}
\item[i)]  the term involving the parameter $B(m_b)$, which also 
appears in the mass difference, is negligible;
\item[ii)] the NLO correction to the coefficient is large and reduces 
the width difference. In LO we obtain the coefficients 0.011 and 0.203
instead of 0.007 and 0.132, respectively;
\item[iii)] the $1/m_b$ correction is also large and negative, and 
its importance is amplified by the negative NLO correction.
\end{itemize}
As a consequence recent estimates of the width difference tend to 
be significantly smaller than the leading order estimate of 
\cite{bbd} and those based on the factorization approximation 
($B=B_S=1$).

There is another way of representing the result (\ref{numform})
\cite{dglait} based 
on the observation that $B(m_b)$ appears in the mass difference an the 
fact the mass difference for $B_d$ mesons is accurately determined
experimentally. Then
\begin{equation}
\label{romeform}
\hspace*{0cm}
\left(\frac{\Delta \Gamma}{\Gamma}\right)_{B_s}\! = 
\left(\!\tau_{B_s}\Delta m_{B_d}\frac{M_{B_s}}{M_{B_d}}\right)
\left|\frac{V_{ts}}{V_{td}}\right|^2 \!K\xi^2 
\left[0.030-0.937 R_S(m_b) - \frac{0.35}{B} \right]\!.
\end{equation}
In this representation the value of the first bracket can be taken 
from experiments, and 
\begin{equation}
K=\frac{4\pi}{3}\frac{m_b^2}{m_W^2}
\left|\frac{V_{cb}V_{cs}}{V_{ts} V_{tb}}\right|^2
\frac{1}{\eta_B(m_b) S_0(x_t)}
\end{equation}
is a known factor, if we assume that the CKM matrix is unitary. The 
advantage of this representation is that hadronic uncertainties 
enter only in ratios, i.e.\ in
\begin{equation}
\xi^2=\frac{f_{B_s}^2 B_{B_s}(m_b)}{f_{B_d}^2 B_{B_d}(m_b)}, 
\quad
R_S(m_b)=-\frac{5}{8} \frac{M_{B_s}^2}{(\bar{m}_b+\bar{m}_s)^2}
\frac{B_S(m_b)}{B(m_b)},
\end{equation} 
and these are believed to be better known than $f_{B_s}$ and 
$B_S(m_b)$. This advantage is more than compensated by 
the need to know $|V_{ts}/V_{td}|$ in Eq.~(\ref{romeform}), 
which makes the prediction for $\Delta\Gamma_{B_s}$ sensitive 
to the global fits to the unitarity triangle and the theoretical 
assumptions that go into it. In particular, the prediction for 
$\Delta\Gamma_{B_s}$ now depends on the assumption that the standard 
model describes flavour mixing correctly, even though the decay 
of $B_s$ meson is unlikely to be affected by new physics in 
mixing. For this reason we would rather discourage the use 
of this method to obtain  $\Delta\Gamma_{B_s}$.

In Table~\ref{tab1} we present a summary of recent estimates 
for $\Delta\Gamma_{B_s}$, based on either Eq.~(\ref{numform}) 
(default) or Eq.~(\ref{romeform}). (We use here only estimates based 
on lattice calculations of  $B(m_b)$ 
and $B_S(m_b)$.) It is evident that there is 
remarkable agreement on the hadronic parameters $B(m_b)$ 
and $B_S(m_b)$, and that the spread of results is mainly 
caused by different values of $f_{B_s}$ or the use of 
Eq.~(\ref{romeform}). In the following we first assess the 
theoretical uncertainties of the NLO calculation and then present 
our best evaluation, given as ``this summary'' in Table~\ref{tab1}. 
The current theoretical uncertainties are as follows:

\begin{table}[t]
\addtolength{\arraycolsep}{0.2cm}
\renewcommand{\arraystretch}{1.3}
$$
\begin{array}{|c||c|c|c|c|}
\hline
& B(m_b) & \bar{B}_S(m_b)  & 
f_{B_s}/\mbox{MeV} & (\Delta\Gamma/\Gamma)_{B_s} \\ 
\hline\hline
\mbox{BBGLN98 \cite{dgbsnlo}} & 
0.90 & 1.07 & 210 & 6.0\%\\  \hline
\mbox{Hashimoto {\em et al.} \cite{dglajap}} & 
0.85\pm 0.11 & 1.24\pm 0.16 & 245  & 10.7\% \\  \hline
\mbox{Becirevic {\em et al.} \cite{dglait}} & 
0.91(3)^{+0.00}_{-0.06} & 1.32(3)^{+0.03}_{-0.04} & 
\mbox{uses Eq.~(\ref{romeform})} & 4.7\%\\  \hline
\mbox{Gimenez/Reyes \cite{dglaes}} & 
0.83\pm 0.08 & 1.25\pm 0.14 & 
\mbox{uses Eq.~(\ref{romeform})} & 5.1\%\\  \hline
&&&&\\[-0.65cm]\hline
\mbox{this summary} & 
0.9\pm 0.1 & 1.25\pm 0.1 & 230\pm 10\%  & 9.3\%\\  \hline
\end{array}
$$
\caption{\label{tab1}
Summary of recent evaluations of $(\Delta\Gamma/\Gamma)_{B_s}$. For 
theoretical uncertainties on   $(\Delta\Gamma/\Gamma)_{B_s}$ consult 
text. $\bar{B}_S(m_b)\equiv M_{B_s}^2 B_S(m_b)/(\bar{m}_b+\bar{m}_s)^2$.}
\end{table}

\begin{itemize}
\item[i)]
{\em Residual scale dependence.} This dependence arises from residual 
effects in the matching of the NLO Wilson coefficients for 
$\Delta \Gamma_{B_s}$ to the parameters $B$ and $B_S$ computed on 
the lattice, and from the matching of the NLO Wilson coefficients for 
$\Delta \Gamma_{B_s}$ to the Wilson coefficients of the $\Delta B=1$ weak 
effective Hamiltonian. Here we estimate only the second source of
scale dependence, using the coefficients for scales $m_b/2$ and 
$2 m_b$ ($m_b=4.8\,$GeV) given in \cite{dgbsnlo}. We then find 
\begin{equation}
\delta\left(\frac{\Delta \Gamma}{\Gamma}\right)_{\rm scale} = 
\left(\begin{array}{c} \!+1.1\%\!\\[-0.1cm]\!-2.7\%\!\end{array}\right)\cdot
\left(\frac{f_{B_s}}{230\,\mbox{MeV}}\right)^{\!2}  
\frac{M_{B_s}^2 B_S(m_b)}{(\bar{m}_b+\bar{m}_s)^2}.
\end{equation}

\item[ii)]
{\em Normalization.} There is an overall normalization error in 
Eq.~(\ref{numform}), which results from the uncertainty in the 
measurement of the semi-leptonic branching fraction and from the 
value of the $b$ quark mass. We estimate this error to be 
$10\%$.

\item[iii)]
{\em $1/m_b$-correction.} The second major source of theoretical 
uncertainty originates from the term ``$-0.078$'' in
Eq.~(\ref{numform}), which has been obtained using the factorization 
assumption for the power-suppressed four-quark matrix elements 
\cite{bbd}. To estimate this uncertainty, we introduce four parameters
that measure the deviation from the factorization approximation,
corresponding to the four independent operators at order $1/m_b$. 
We find that all $1/m_b$ effects add essentially constructively, 
so that there is no reason to suspect an amplification of corrections 
to the factorization approximation as a result of cancellations in the
factorized expression. We then find 
\begin{equation}
\delta\left(\frac{\Delta \Gamma}{\Gamma}\right)_{1/m_b} = 
\left(\pm 6\%\,\Delta r\right)\cdot
\left(\frac{f_{B_s}}{230\,\mbox{MeV}}\right)^{\!2},
\end{equation}
where $\Delta r$ parametrizes a typical deviation from the
factorization approximation, corresponding to quantities like 
$B-1$, and $B_S-1$ in leading power. There exists no reliable
information on violations of the factorization assumption for the 
matrix elements in question, but experience with leading order matrix
elements suggests that $\Delta r\approx 0.3$ is a reasonable upper 
limit. 

\item[iv)]
{\em $B(m_b)$ and $B_S(m_b)$.} Substantial work has been invested in 
controlling the non\-per\-tur\-bative parameters at leading
power. Table~\ref{tab1} shows that the results are rather 
consistent with each other (the number 1.07 in the first line 
is obsolete) and suggests that 
\begin{equation}
B(m_b)=0.9\pm 0.1, \qquad \bar{B}_S(m_b)\equiv 
\frac{M_{B_s}^2 B_S(m_b)}{(\bar{m}_b+\bar{m}_s)^2}=1.25\pm 0.10. 
\end{equation}
We note that at the present stage the uncertainty in $B$ and $B_S$  
has become a minor factor, the dominant ones coming from residual 
scale dependence and $1/m_b$ corrections.
\end{itemize}

\noindent
We can now combine these uncertainties with Eq.~(\ref{numform}) to
obtain our final result
\begin{eqnarray}
\label{update}
\hspace*{0cm}
\left(\frac{\Delta \Gamma}{\Gamma}\right)_{B_s} &=& 
\left(\frac{f_{B_s}}{230\,\mbox{MeV}}\right)^{\!2} \Big(1\pm 0.1\Big)
\Big[(13.2^{+1.1}_{-2.7})\% \,\bar{B}_S(m_b) +0.7\%\,B(m_b) 
\nonumber\\[-0.1cm] 
&&\hspace*{4.3cm} -\, (7.8\pm 1.8\%)\Big]
\nonumber\\[-0.0cm]
&=& \Big(9.3^{+3.4}_{-4.6}\Big)\%.
\end{eqnarray}
The final number is obtained by adding all errors in squares and
assigning a $10\%$ uncertainty to $f_{B_s}=230\,$MeV. It is worth 
noting that this number is considerably larger than the result obtained 
in \cite{dglait,dglaes} for two reasons: first, the overall 
normalization obtained from Eq.~(\ref{romeform}) is about 25\% smaller 
compared to that of Eq.~(\ref{numform}) with $f_{B_s}=230\,$MeV, 
since the global CKM fit prefers a smaller value of the $B$ meson 
decay constant; second, Ref.~\cite{dglait} uses $m_b=4.6\,$GeV, 
which increases the $1/m_b$ correction by almost a factor 1.5. 

\section{Prospects for improvement} 

Despite (or because of?) extensive work on radiative and $1/m_b$ 
corrections the theoreti\-cal prediction of the width difference of the 
$B_s$ mass eigenstates remains rather uncertain. This is due to an 
unfortunate conspiracy of negative corrections at next-to-leading 
order in $\alpha_s$ and in the heavy quark expansion. The large error 
exhibited by Eq.~(\ref{update}) makes it improbable that new physics 
would be first observed in the width difference, since a new physics 
contribution to the $B_s$-$\bar{B}_s$ mixing phase large enough 
to cause an effect larger than the theoretical uncertainty would 
rather been seen elsewhere, for instance as a time-dependent 
asymmetry in $B_s\to J/\psi \phi$.

Is further improvement possible? It is conceivable that future lattice
calculations will reduce further the uncertainty in $B$ and $B_S$, 
but the impact of the present uncertainty in these parameters on the 
total error is no longer dominant. Some improvement could be possible 
concerning the overall normalisation, but this effect can also not be 
substantial. The obvious targets for improvement are therefore

\begin{itemize}

\item[ii)] scale dependence: its reduction would require the 
calculation or estimate of $\alpha_s^2$ corrections. This appears 
to be a hard endeavour, and it would also require the computation 
of three-loop anomalous dimensions of the weak effective 
Hamiltonian.

\item[ii)] $1/m_b$ corrections: obviously, the relevant matrix 
elements should be computed on the lattice to rid the calculation 
from the factorization assumption. Such a calculation does not 
appear likely soon, given the familiar difficulties with higher 
dimension operators. Since the $1/m_b$ corrections are large, one 
may think of computing the $\alpha_s$ corrections to them. While this 
is feasible, the effort is probably too large as long as the 
matrix elements of the operators are not known accurately enough.

\end{itemize}

\noindent
It therefore appears that $(\Delta \Gamma/\Gamma)_{B_s}$ will remain 
considerably uncertain for the foreseeable future. 

\ack
We wish to thank the organizers of the workshop for their successful 
work, and G. Buchalla, C. Greub and U. Nierste for collaboration.

\end{document}